\documentclass[useAMS]{mn2e}

\usepackage{graphicx,color,here}
\usepackage{amssymb}
\usepackage{amsmath}

\author[R.~P. Eatough et al.]{R.~P. Eatough$^1$\thanks{E-mail: ralph.eatough@pulsarastronomy.net}, E.~F. Keane$^1$, A.~G. Lyne$^1$\\ $^1$ Jodrell Bank Centre for Astrophysics, Alan Turing Building, University of Manchester, Manchester, M13 9PL, United Kingdom.} \date{1 May 2008}
\title{An Interference Removal Technique \\ for Radio Pulsar Searches}

\begin{document}

\maketitle

\vspace{3cm}
\begin{abstract}
Searches for radio pulsars are becoming increasingly difficult because
of a rise in impulsive man-made terrestrial radio-frequency
interference. Here we present a new technique, zero-DM filtering,
which can significantly reduce the effects of such signals in pulsar
search data. The technique has already been applied to a small portion of
the data from the Parkes multi-beam pulsar survey, resulting in the
discovery of four new pulsars, so illustrating its efficacy.
\end{abstract}


\vspace{3cm}
\pagenumbering{arabic}
\pagestyle{plain}

\section{Introduction}
Pulsar search data are often corrupted by the presence of impulsive,
broadband and sometimes periodic terrestrial radiation. Such {\it
radio frequency interference} (RFI) originates in unshielded
electrical equipment which produces discharges, such as automobile
ignition systems, electric motors, and fluorescent lighting, as well
as discharge from high-voltage power transmission lines.  It may also
arise naturally from the radio emission generated in lightning
discharges. Such RFI can often be very strong and can even enter
receiver systems through the far-out sidelobes of the telescope
reception pattern. Several methods have been employed to reduce the
effects of RFI, such as clipping intense spikes, removing parts of the
fluctuation spectrum, and identifying common signals in the different
beams of a multi-beam receiver system. These procedures all require
carefully tuned algorithms to remove the interference, and at the same
time must cause minimal damage to the astronomical data.

RFI signals are mostly broadband but generally do not display the
dispersed signature of radio pulsars. Here, we present a simple
algorithm which we call `zero-DM filtering' to selectively remove
broadband undispersed signals from data prior to the application
of normal pulsar search algorithms. The outline of this paper is as
follows: following the description of the algorithm in section 2, in
section 3 we show that the procedure does indeed remove much of the
interference in typical observing situations and in general greatly
improves the visibility of celestial dispersed pulses. In section 4 we
present four new pulsars discovered in the Parkes multi-beam pulsar
survey (PMPS) after applying the new technique. Finally section 5 is a
discussion of the benefits and limitations of the procedure.

\section{Zero-DM Filtering}
A fundamental characteristic of broadband pulsar signals is the
frequency-dependent dispersion of their pulse arrival times as a
result of traversing the ionised component of the interstellar
medium. Pulses observed at lower frequency arrive later than their
higher-frequency counterparts.  The delay, $t$, at radio frequency $f$ (MHz)
is given by:
\begin{equation}
\label{eq:disp_t}
 {t}=4150{\left(\frac{\rm DM}{f^{2}}\right)\;\; {\rm sec,}}
\end{equation}
where DM is the {\em dispersion measure}, the integrated column
density of free electrons along the line of sight, measured in
cm$^{-3}$pc (e.g. Lyne \& Smith 2006)\nocite{ls06}. Detecting a pulse
with a finite bandwidth receiver results in a broadened pulse profile
and a corresponding reduction in pulse signal-to-noise ratio.  Before
any search for pulses can be performed, the data have to be
compensated for the effects of a number of trial values of DM.  
In order to do
this, the bandwidth of the receiver is first split into a number of
independent frequency channels, which are
appropriately sampled to produce a two-dimensional array of samples,
${\rm S}(f_{\rm i},t_{\rm j})$ at frequency $f_{\rm i}$ and time
$t_{\rm j}$. Appropriate time delays for a given trial DM are then
applied to each frequency channel so that any pulses at this DM are
aligned in time.  The frequency channels are then summed together to
produce a time series of dedispersed data, $x(t)$.

\begin{figure}
\begin{center}
\includegraphics[scale=0.3]{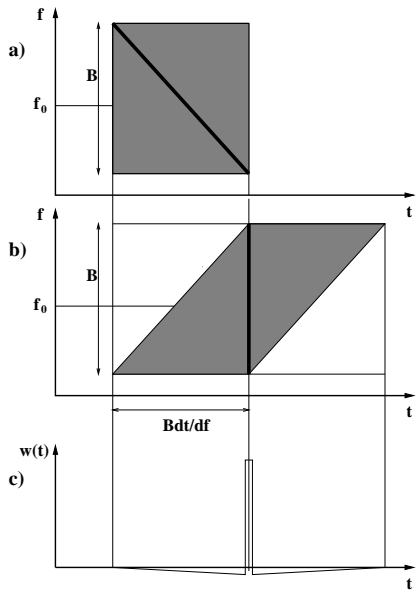}
\caption{The effect of the the zero-DM filter on an idealised narrow
linearly-dispersed pulse in the frequency-time domain. The top panel
(a) shows the dispersion drift of the pulse ${\rm B}\;{\rm d}t/{\rm d}f$
over the full bandwidth B of the receiver.  Subtraction of the mean in
vertical strips to remove zero-DM signals (Equation~\ref{eq:zero-DM})
leaves the non-pulse area (shaded grey) negative. Panel (b) shows the
result of dedispersing the data at the correct value of DM and the
lower panel (c) shows the resultant pulse shape after adding all the
frequency channels together.}
\label{fig:zdm}
\end{center}
\end{figure}

The zero-DM filter is implemented prior to the dedispersion described
above, by simply calculating the mean of
all frequency channels in each time sample and subtracting this from
each individual frequency channel in the time sample. Hence the
adjusted values of the samples, ${\rm S'}(f_{\rm i},t_{\rm j})$ are
calculated as:
\begin{equation}
{{\rm S'}}(f_{\rm i},t_{\rm j}) = {{\rm S}}(f_{\rm i},t_{\rm j}) - {1 \over n_{\rm chans}} 
\sum_{i=1}^{n_{\rm chans}} {\rm S}(f_{\rm i},t_{\rm j}). 
\label{eq:zero-DM}
\end{equation}

\begin{figure}
\begin{center}
\includegraphics[scale=0.3, angle=-90]{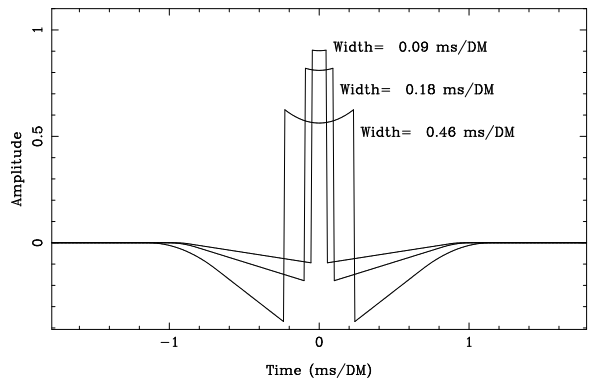}
\caption{The pulse distortion caused by zero-DM filter on idealised
box-car pulses of varying width for the PMPS observing system. The
time scale should be multiplied by the DM in cm$^{-3}$pc to give the
time in milliseconds.}
\label{fig:distn}
\end{center}
\end{figure}

Clearly, any broadband undispersed signal will result in a
simultaneous rise across all frequency channels and will be removed by
this process.  We now investigate the effect of this procedure on
dispersed cosmic pulses.  Consider an idealised dispersed pulsar
signal with very narrow pulses and approximate the dispersion drift
across the full receiver bandwidth ${\rm B}$ to a linear slope,
${\rm d}t/{\rm d}f$, which is given approximately by:
\begin{equation}
\frac{{\rm d}t}{{\rm d}f}=-8300 \left( \frac{\rm DM}{{f_{\rm 0}}^3} \right) \;\;{\rm sec/MHz,}
\end{equation}
where $f_{\rm 0}$ (MHz) is the central observing frequency.
Figure~\ref{fig:zdm}a shows such a pulse sweeping through the
frequency band of a receiver, traversing the full bandwidth in a time
of ${\rm B}\; {\rm d}t/{\rm d}f$.  After subtraction of the mean in
vertical strips as described by Equation~\ref{eq:zero-DM}, the
non-pulse grey area in the Figure~\ref{fig:zdm}a is
negative. Dedispersion then distorts this pattern in the manner shown
in Figure~\ref{fig:zdm}b, so that adding vertically after the
dedispersion then gives the convolving function, $w(t)$ shown in
Figure~\ref{fig:zdm}c.  Note the negative triangular area which is
adjacent to, and equals the area of, the pulse.  Thus the zero-DM
process results in the dedispersed time series $x(t)$ being convolved
by this function, $w(t)$. 
Of course any real celestial pulse will have a finite width. In
Figure~\ref{fig:distn} we show the effects of the zero-DM filter on
idealised box-car pulses of different widths for the PMPS observing
system at 1.4 GHz (Manchester et al. 2001)\nocite{mlc+01}.  In this
case, $f_{\rm 0}=1394$~MHz and B=288~MHz. The pulse width in
milliseconds can be calculated by simply multiplying by the DM in
cm$^{-3}$pc. For example, at a DM of 100 cm$^{-3}$pc the bottom pulse
would have a width of 46 ms. The peak amplitude clearly decreases for
pulses with larger widths.

\begin{figure}
\begin{center}
\includegraphics[scale=0.3, angle=-90]{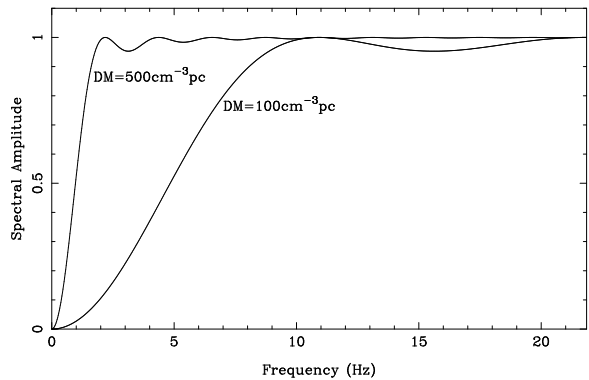}
\caption{\small Two example effective zero-DM filters for
DM=500~cm$^{-3}$pc and DM=100~cm$^{-3}$pc for the PMPS, showing
variation in relative spectral amplitude with the fluctuation
frequency.}
\label{fig:spec}
\end{center}
\end{figure}

In the frequency domain, equivalently, the fluctuation spectrum of the
dedispersed times series is multiplied by the Fourier
transform, $W(\nu)$, of $w(t)$ which can be shown to be of the
form:
\begin{equation}
\begin{split}
W(\nu) & = 1-{\rm sinc}^2(\pi {\rm B}\frac{{\rm d}t}{{\rm d}f} \nu) \\
 & = 1-{\rm sinc}^2 \left( \frac{8300\pi{{\rm B\;DM}}}{f_{\rm 0}^3} \nu \right),
\label{eq:spec}
\end{split}
\end{equation} 
where $\nu$ is the fluctuation frequency of the signal. 
Figure~\ref{fig:spec} shows the effect of the zero-DM filter on the
spectral amplitude of a pulse 
at two specified DM's for PMPS-style observations. 
The zero-DM process effectively acts as a high-pass filter
on the dispersed pulse.
The `low-frequency cutoff', $\nu_{\rm cutoff}$ can be shown using
Equation~\ref{eq:spec} to lie at a fluctuation frequency $\sim$500/DM
Hz. Thus for DM=100 cm$^{-3}$pc, all frequencies $<$5~Hz will be lost.
The reduction in amplitude and pulse distortion we see in
Figure~\ref{fig:distn} can be explained by the loss of the low
frequency components of the pulse profile.

For a typical pulsar we measure a regular train of narrow pulses,
the spectrum of which is a `picket fence.' The zero-DM filter will
cut off all spectral features below $\nu_{\rm cutoff}$.
However, normal pulsars with pulse widths a few percent of their
pulse period will usually have most of
their power in spectral harmonics at higher frequencies and these will
not be affected significantly.  Since standard pulsar searches perform
harmonic summing (see e.g. Lyne \& Smith 2006 or, for a detailed
description of pulsar search algorithms, Lorimer \& Kramer
2005\nocite{lk05}),\nocite{ls06} the detrimental effect of the zero-DM
filter on the final spectral amplitude is reduced.  Assuming the
pulsar has $n\sim P/{2W}$ harmonics of equal amplitude where $P$ is
the pulsar period and $W$ is the pulse width, the number of
harmonics lost below the low-frequency cutoff is given by $n_{\rm
cutoff}\sim P \nu_{\rm cutoff}$, the harmonic summing procedure
will increase the spectral signal-to-noise ratio by a factor of
approximately $(n - {n_{\rm cutoff}})^{1/2}$.



The zero-DM process causes a predictable reduction in the amplitudes
of low frequencies of the fluctuation spectrum of a dispersed pulsar.
It will also reduce the amplitude of any broadband undispersed
impulsive signals substantially, but will have little effect upon
thermal noise.  The signal-to-noise ratio of the pulsar can therefore
be optimised by applying an optimum matched filter to the spectrum,
which reduces the spectral amplitude where the pulsar signal is known
to be small because of the zero-DM filtering.  Thus, there is a second
step to the filtering process whereby the frequency response of the
zero-DM filter is re-applied after each dedispersion trial.  In the
time domain this is best illustrated by considering a single-pulse
search (McLaughlin et al. 2006)\nocite{mll+06}.  Standard single-pulse
searches involve convolution of the dedispersed time series $x(t)$
with box-cars $b(t)$ of various widths and searching for peaks in
$x(t)*b(t)$, where $*$ denotes convolution.  In our case, the zero-DM
filtered time series is $x'(t)=x(t)*w(t)$, so that a standard
single-pulse search looks for peaks in $x'(t)*b(t)$. As a result of
zero-DM filtering the optimal filter is now not a box-car but rather
is given by $b'(t)=b(t)*w(t)$. We now search for peaks in
$x''(t)=x'(t)*b'(t)={\cal F}^{-1}[X'(\nu)*W(\nu)]*b(t)$ where ${\cal
F}$ denotes the Fourier transformation and we have used the
commutability of convolution so the convolution can be performed in
the frequency domain, which is what is done practice.

\section{Examples}
\begin{figure}
\begin{center}
\includegraphics[angle=-90, scale=0.27]{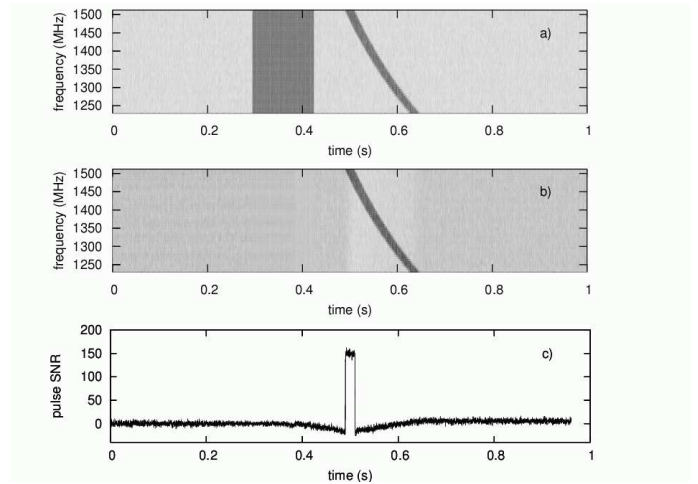}
\caption{\small The top panel (a) shows a grey plot of simulated data
of a 130-ms burst of broadband RFI with DM=0~cm$^{-3}$pc followed by a
20-ms dispersed pulse with DM=150~cm$^{-3}$pc across a 288MHz bandpass
centered at 1374 MHz. The middle panel (b) shows the same data after
application of the zero-DM filter. Note the negative non-pulse
area. In panel (c), the data have been dedispersed for a
DM=150~cm$^{-3}$pc and summed in frequency giving the pulse profile.}
\label{fig:simul}
\end{center}
\end{figure}

The effects of the zero-DM filter have been investigated using
simulated PMPS data generated with the {\it
Sigproc-4.2}\footnote{\footnotesize http://sigproc.sourceforge.net}
pulsar analysis software suite. Figure~\ref{fig:simul}a shows
simulated data of a 130-ms burst of broadband RFI with DM=0~cm$^{-3}$pc
followed by a 20-ms dispersed pulse with DM=150~cm$^{-3}$pc across a
288MHz bandpass centered at 1374 MHz. Figure~\ref{fig:simul}b shows
the same data after application of the zero-DM filter. The broadband
RFI signal has been completely removed with little effect on the
dispersed pulse. Dedispersion at a DM=150~cm$^{-3}$pc and summation in
frequency produces the pulse profile shown in
Figure~\ref{fig:simul}c. The expected pulse distortion is clearly
visible.

We also present examples of the zero-DM filter as applied to
observational data from the PMPS in both periodicity and single-pulse
searches. All clipping and zapping algorithms used previously for
mitigating the effects of RFI (Hobbs et al. 2004)\nocite{hfs+04} have
been omitted. In our periodicity search examples re-application of the
zero-DM filter after each dispersion trial has not been performed
because of the additional computational overheads this would impose in 
our search algorithms (see Section 5 for a further discussion on this subject).

\begin{figure}
\begin{center}
\includegraphics[angle=-90, scale=0.16]{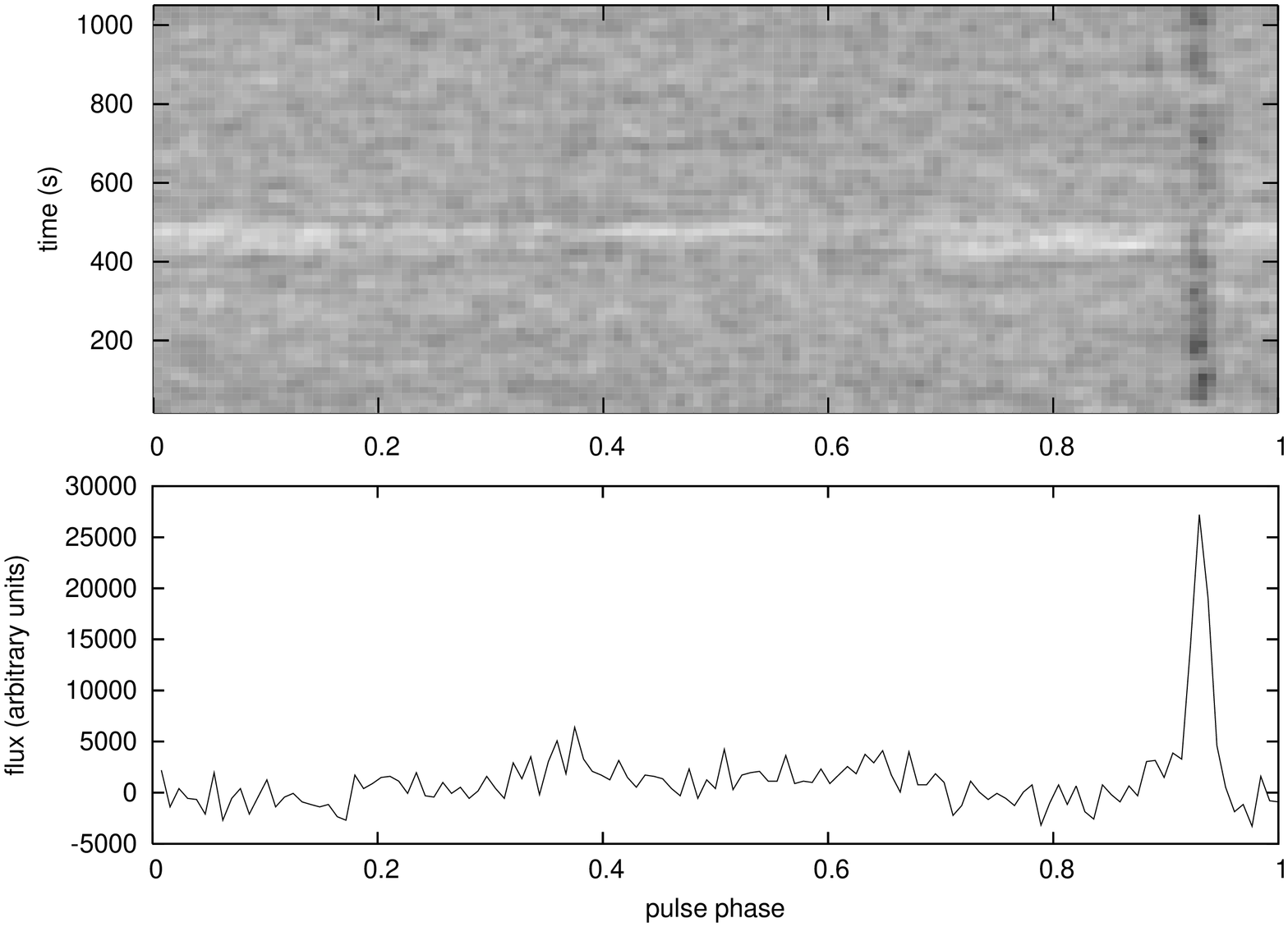}
\includegraphics[angle=-90, scale=0.16]{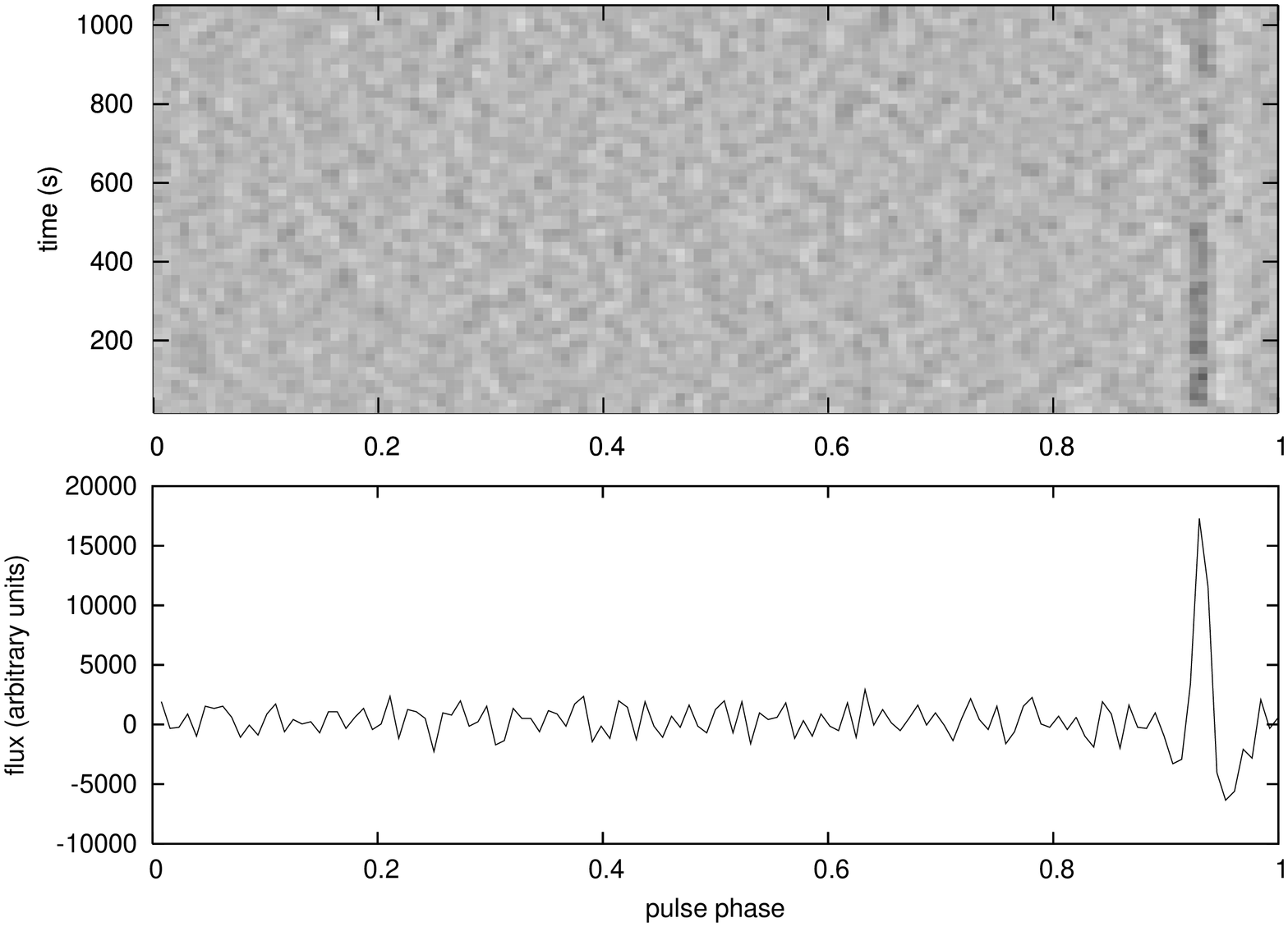}
\caption{\small The effects of the zero-DM filter on folding PMPS data
from the 3.1-second pulsar, PSR~J1842+0257, which has a DM of 148
cm$^{-3}$pc. The left-hand panels are the result of using a regular
analysis, while the right-hand panels are after application of the
zero-DM filter.  On each side, at the top are grey plots of 63
temporal sub-integrations folded at the pulse period, covering the
full 35-minute survey observation, and at the bottom is a panel
showing the full integrated profile. The effect of the filter on the
shape of the pulse profile can be clearly seen. The asymmetry of the
depressions either side of the pulse arise because the dispersion
sweep across the bandpass is not perfectly linear but has a quadratic
component, although in most cases the sweep is well approximated by a
linear slope. Note that the impulsive RFI present through the majority
of the observation is almost completely removed and the baseline
wobble in the final profile is much flatter in the filtered case.}
\label{fig:J1842}
\end{center}
\end{figure}

\begin{figure}
\begin{center}
\includegraphics[scale=0.2]{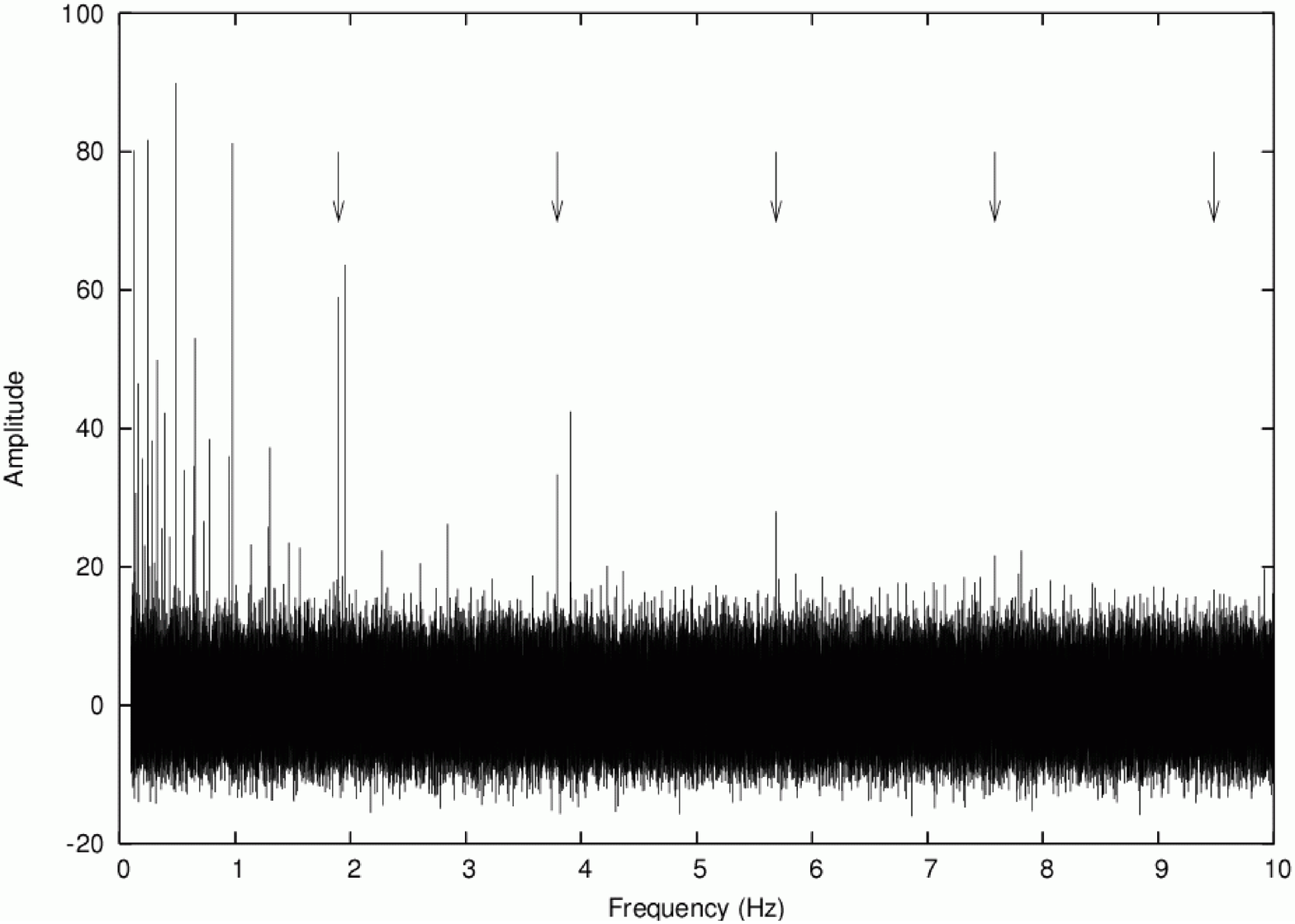}
\includegraphics[scale=0.2]{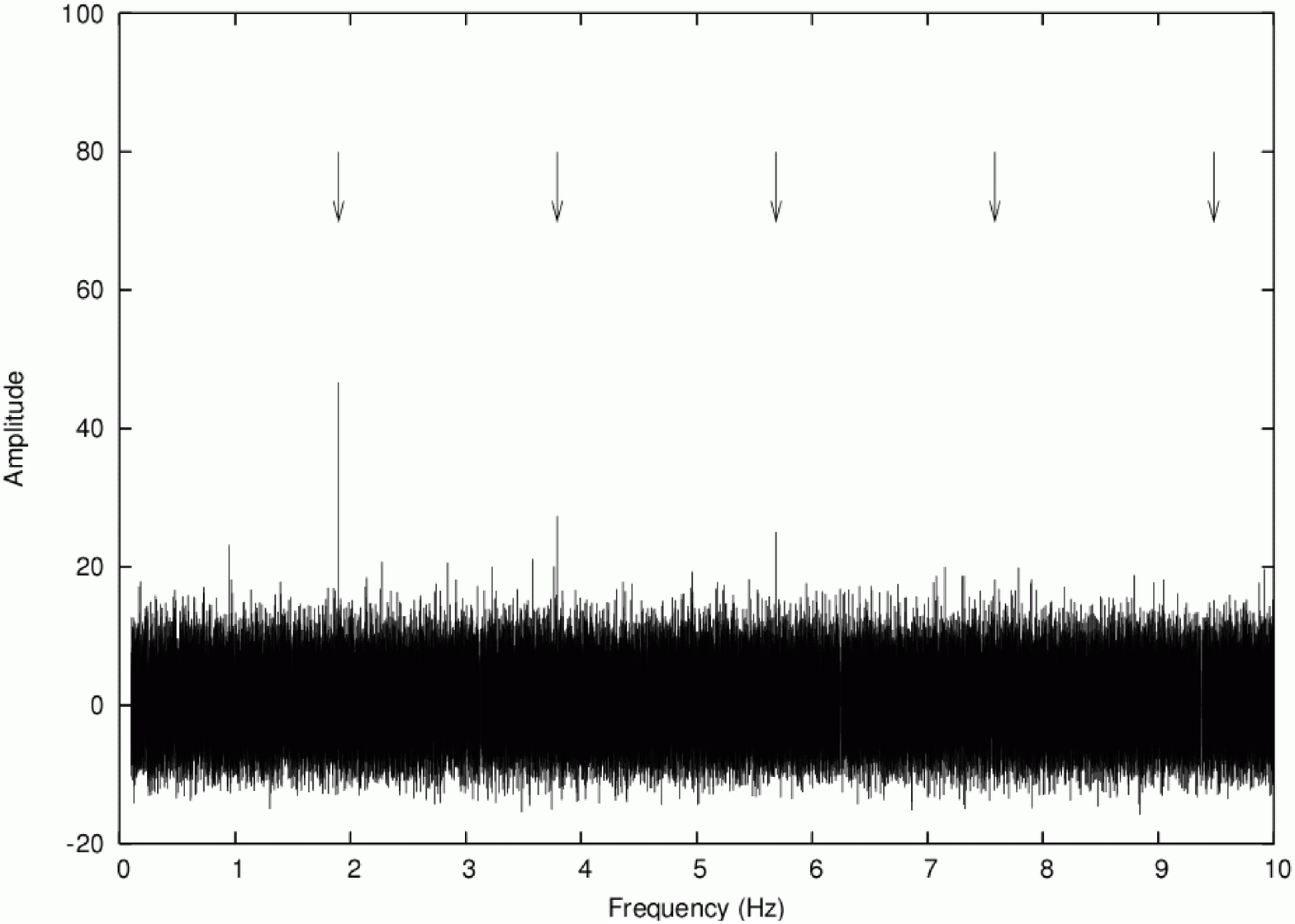}
\caption{\small Fluctuation spectra from a PMPS survey observation in
the direction of the 0.53-second pulsar, PSR~J1604-4718, which has a
DM of 52~cm$^{-3}$pc. The top panel shows the spectrum without any zero-DM
filtering, the bottom panel is the same data after the zero-DM filter
has been applied. The arrows indicate the position of the fundamental
spin frequency of the pulsar and four subsequent harmonics. Most of
the RFI signals are completely removed by the new technique.}
\label{fig:Spectra}
\end{center}
\end{figure}

Figure~\ref{fig:J1842} illustrates clearly the effects of the
procedure on a 35-minute PMPS observation in the direction of
PSR~J1842+0257, which has a period of 3.1 seconds and a DM of 148
cm$^{-3}$pc, so that the zero-DM low-frequency cutoff is at about 3.3
Hz. On the left is the normal search output, while on the right is
that with the zero-DM filtering procedure in place. After application
of the zero-DM filter the red noise and impulsive interference is
almost completely removed, and the expected pulse distortion due to
the filtering as shown in Figure~\ref{fig:zdm}c and
Figure~\ref{fig:simul}c is clear. Remarkably, even with such a
long-period pulsar, in which the fundamental and first few harmonics
are completely missing, the absence of the interference makes its
detection more secure than previously.
A number of periodicity searches on PMPS beams containing known
pulsars with a range of periods and dispersion measures has been
performed. The fluctuation spectra from the periodicity search are
much cleaner than before (see Figure~\ref{fig:Spectra}).  The
`forest' of RFI signals at low frequency are completely removed by
the new technique. The pulsar spectral signal-to-noise ratios behave
as expected, but because of the reduction in the number of periodic
RFI signals, the pulsars are easier to detect. 
 
\begin{figure}
\begin{center}
\includegraphics[scale=0.24]{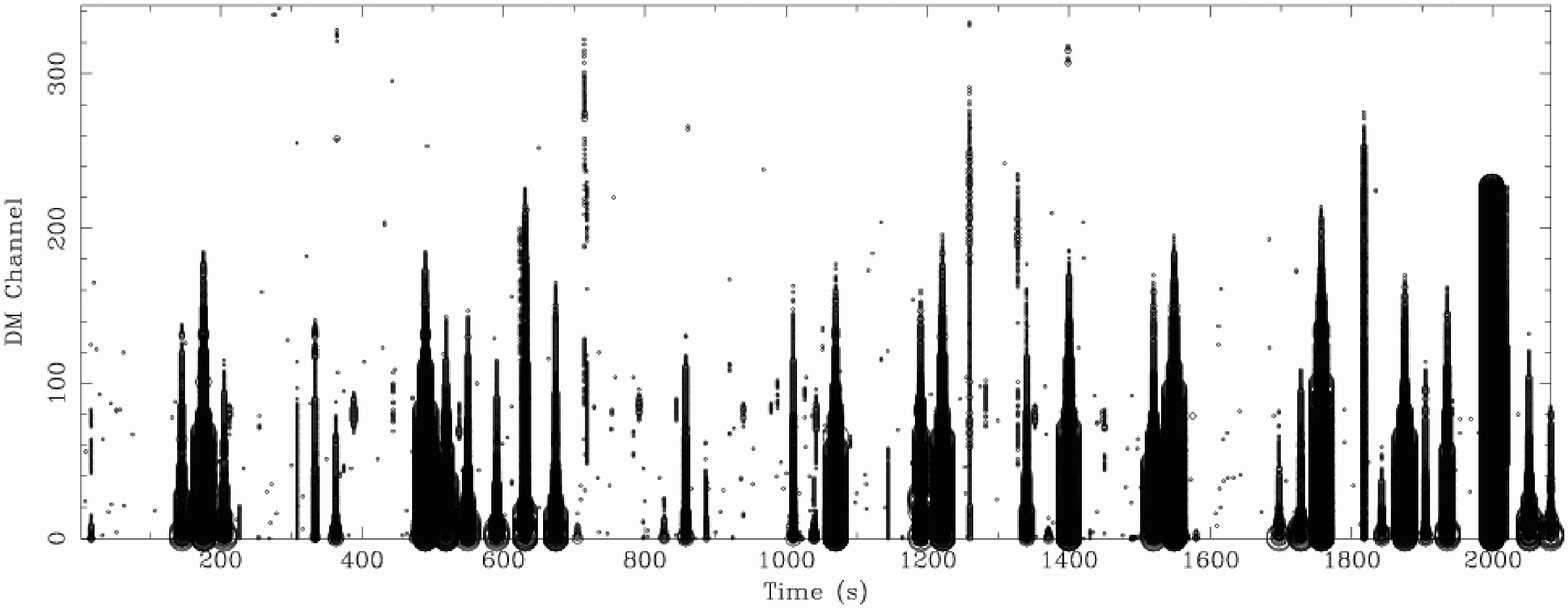}
\includegraphics[scale=0.24]{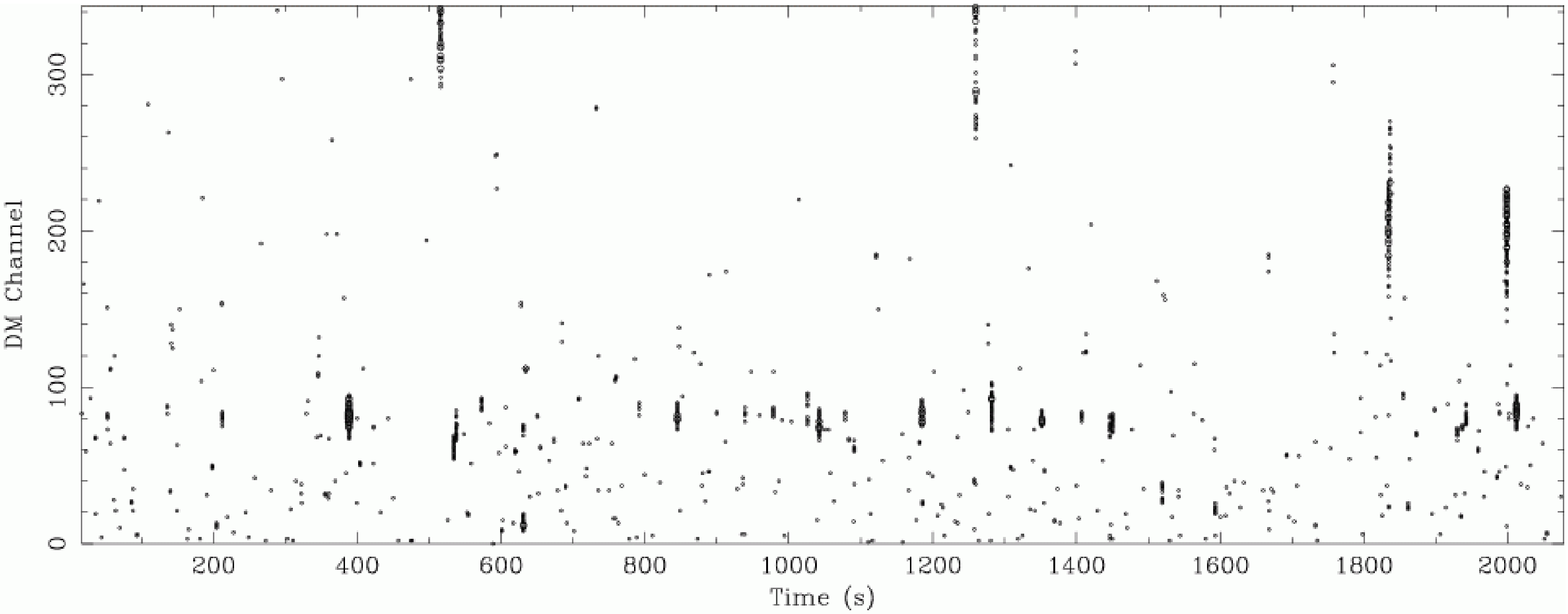}
\includegraphics[scale=0.24]{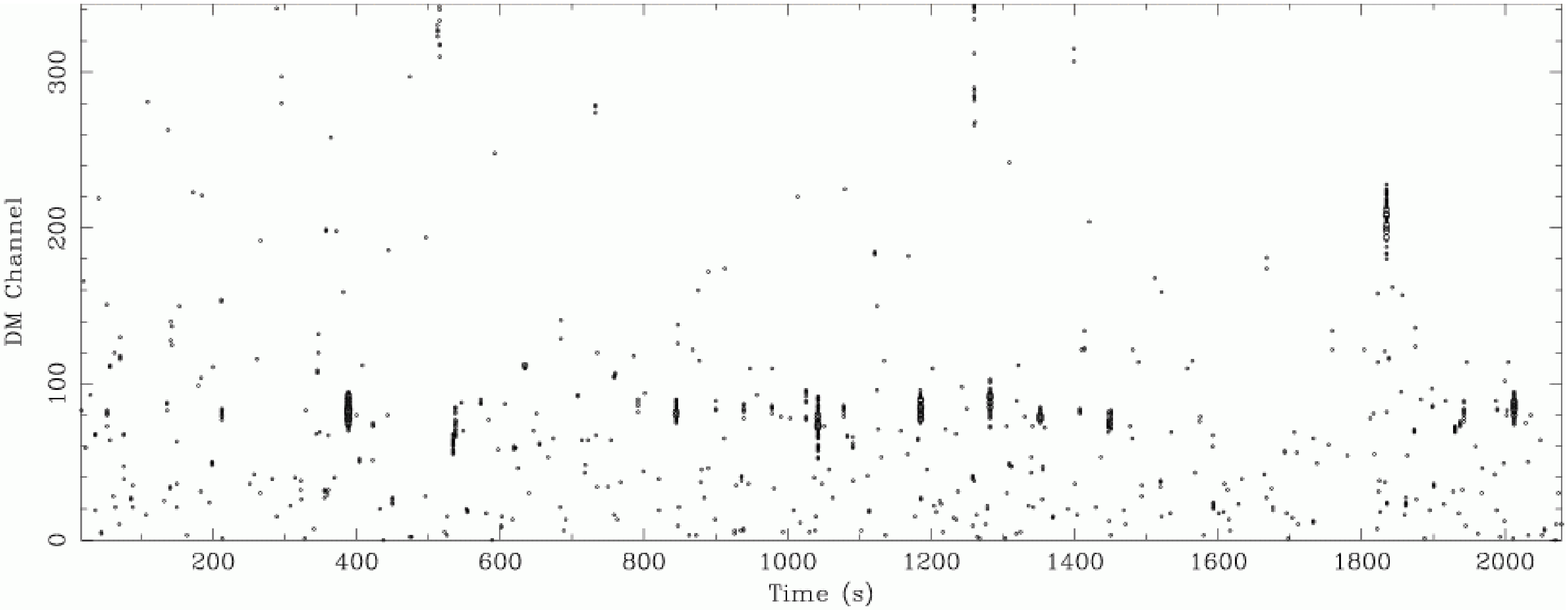}
\caption{\small{Single-pulse search diagnostic plots of a 35-minute
observation of the erratic pulsar PSR~B1735$-$32. Top: using standard
search method without zero-DM filter; Broadband terrestrial
interference causes the large stripes across a wide range of DM.
Middle: After using zero-DM filter, most RFI streaks have been removed
with single pulses from the pulsar now much more visible (at DM
channel$\approx$82). Bottom: After using the filter and applying the
optimal matched filter. Remnant RFI streaks are further removed and
pulses from the pulsar remain.}}
\label{fig:big_plots}
\end{center}
\end{figure}

Figure~\ref{fig:big_plots} shows examples of single-pulse
search diagnostic plots from our analysis of the PMPS. Plotted in
Figure~\ref{fig:big_plots} are the time series for each of the 325
trial DM's (in the range $0-2200$ cm$^{-3}$pc). Each detected
single-pulse event is plotted as a circle with radius proportional to
signal-to-noise ratio. The top panel is the result of a standard
analysis without zero-DM filtering. The observation contains single
pulses from PSR~B1735$-$32 at DM channel=82 (DM=50~cm$^{-3}$pc), but
we can see that the output is contaminated with many RFI streaks
making identification of the pulsar difficult. The middle panel shows
the corresponding plot with zero-DM filtering applied. The vast
majority of the RFI has been removed but the pulsar is still
present. Obvious also are a few remnant RFI streaks at high DM which
have not been removed by the filtering. Searching the time series with
the optimal single-pulse profile, $w(t)$, improves things even further
by removing the remnant streaks almost completely, as can be seen in
the lower panel.

\section{Discovery of Four new Sources}
\begin{figure}
\begin{center}
\includegraphics[angle=-90, scale=0.16]{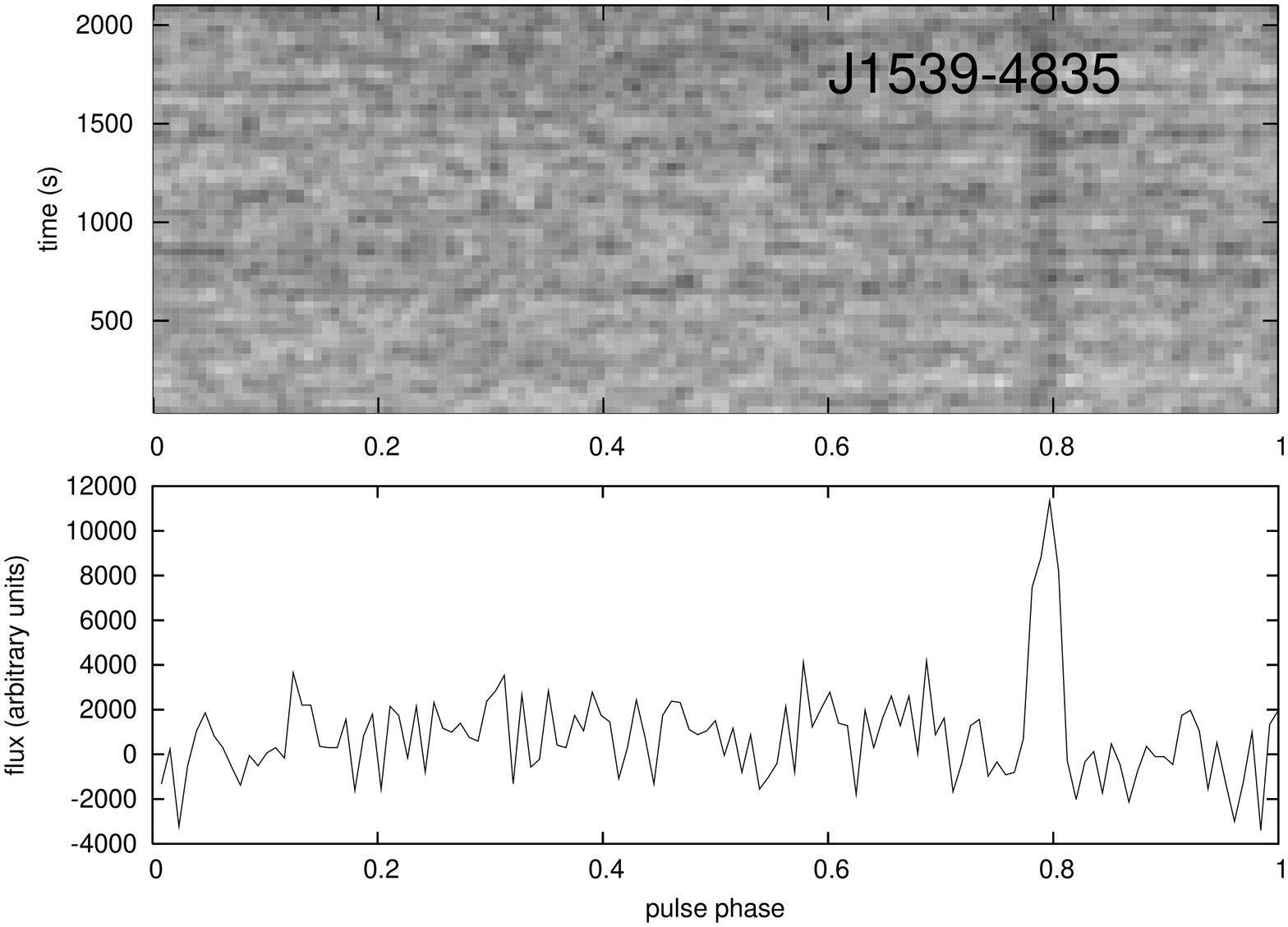}
\vspace{0.45cm}
\includegraphics[angle=-90, scale=0.16]{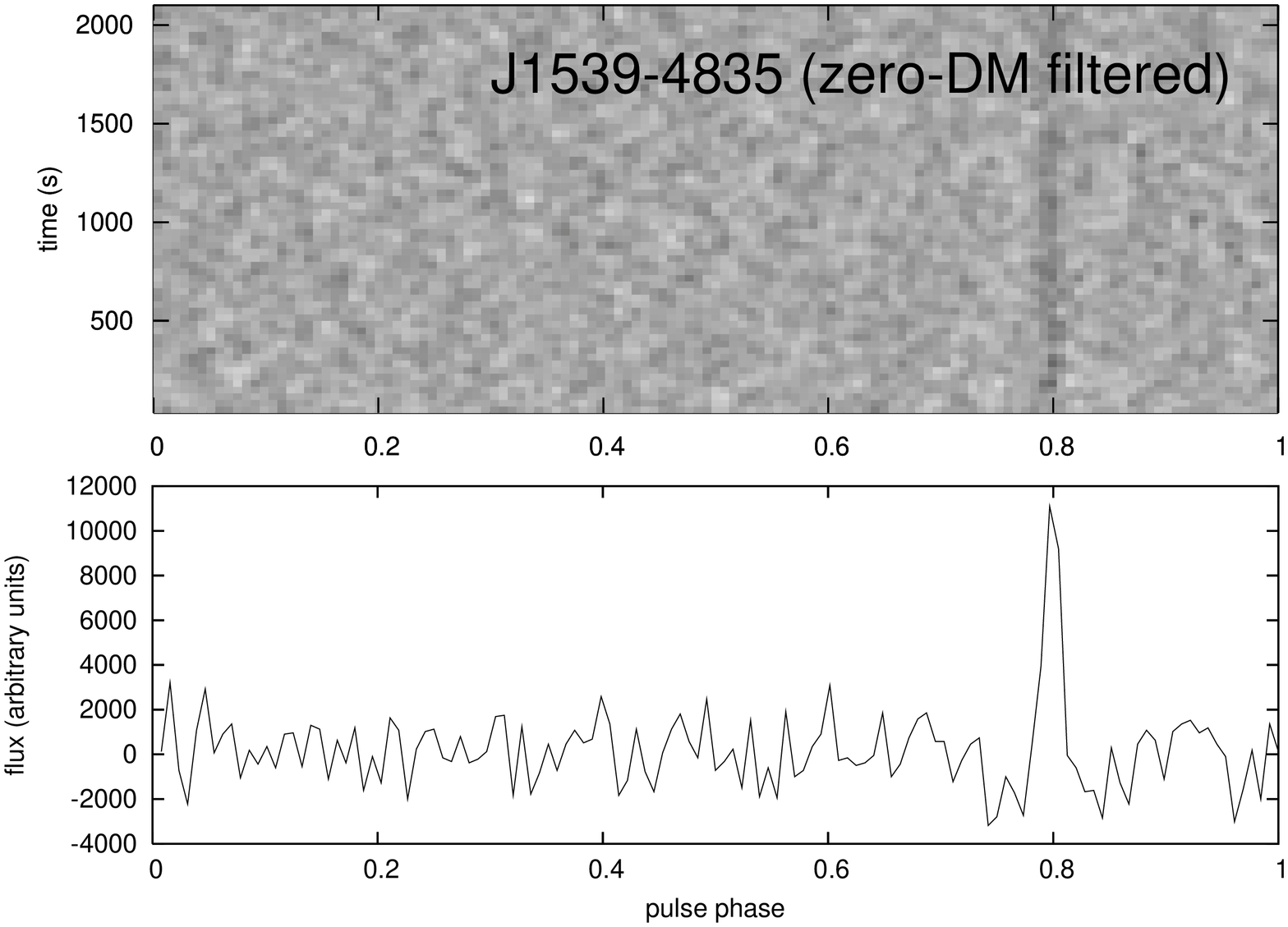}

\includegraphics[angle=-90, scale=0.16]{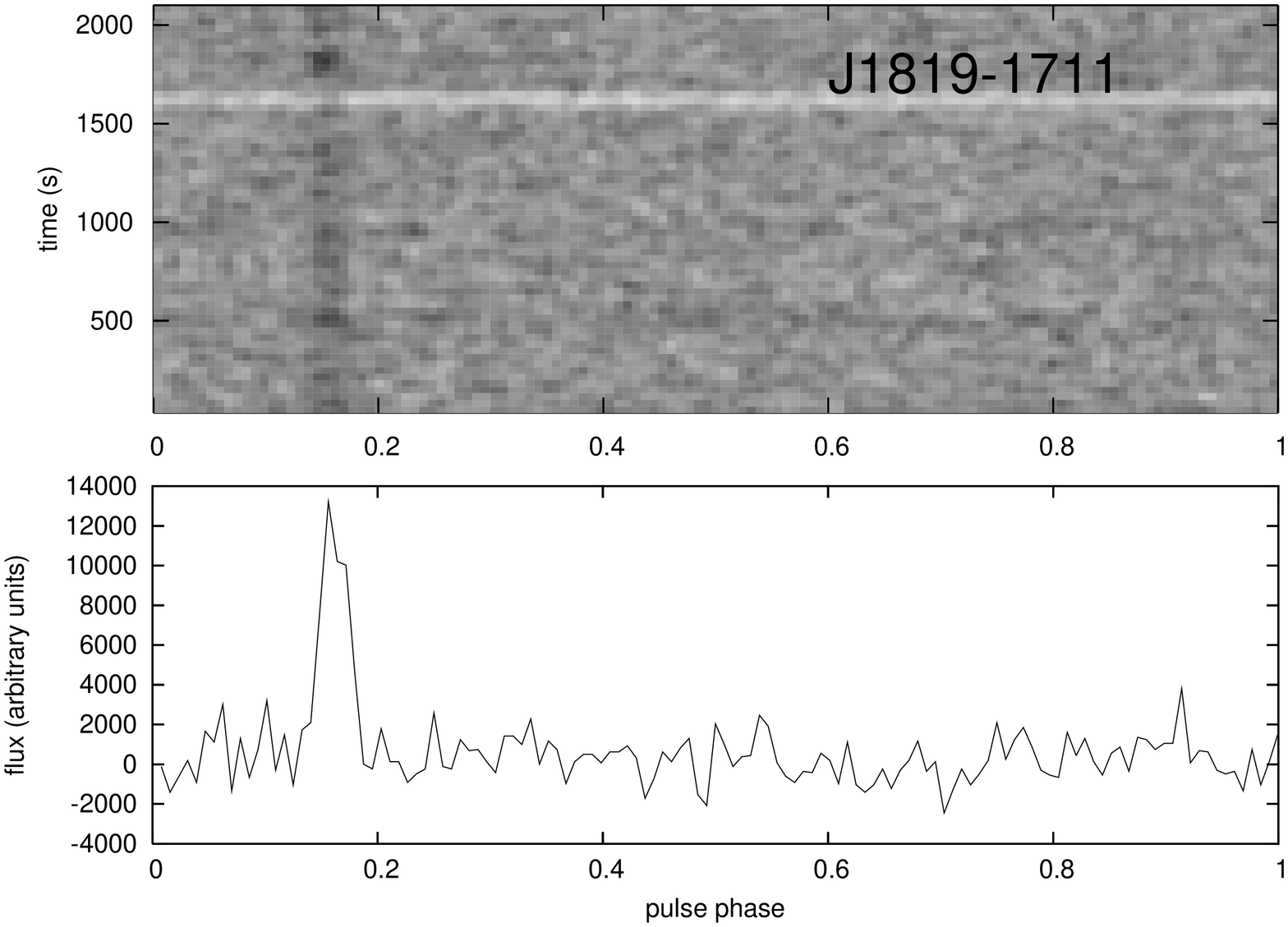}
\vspace{0.45cm}
\includegraphics[angle=-90, scale=0.16]{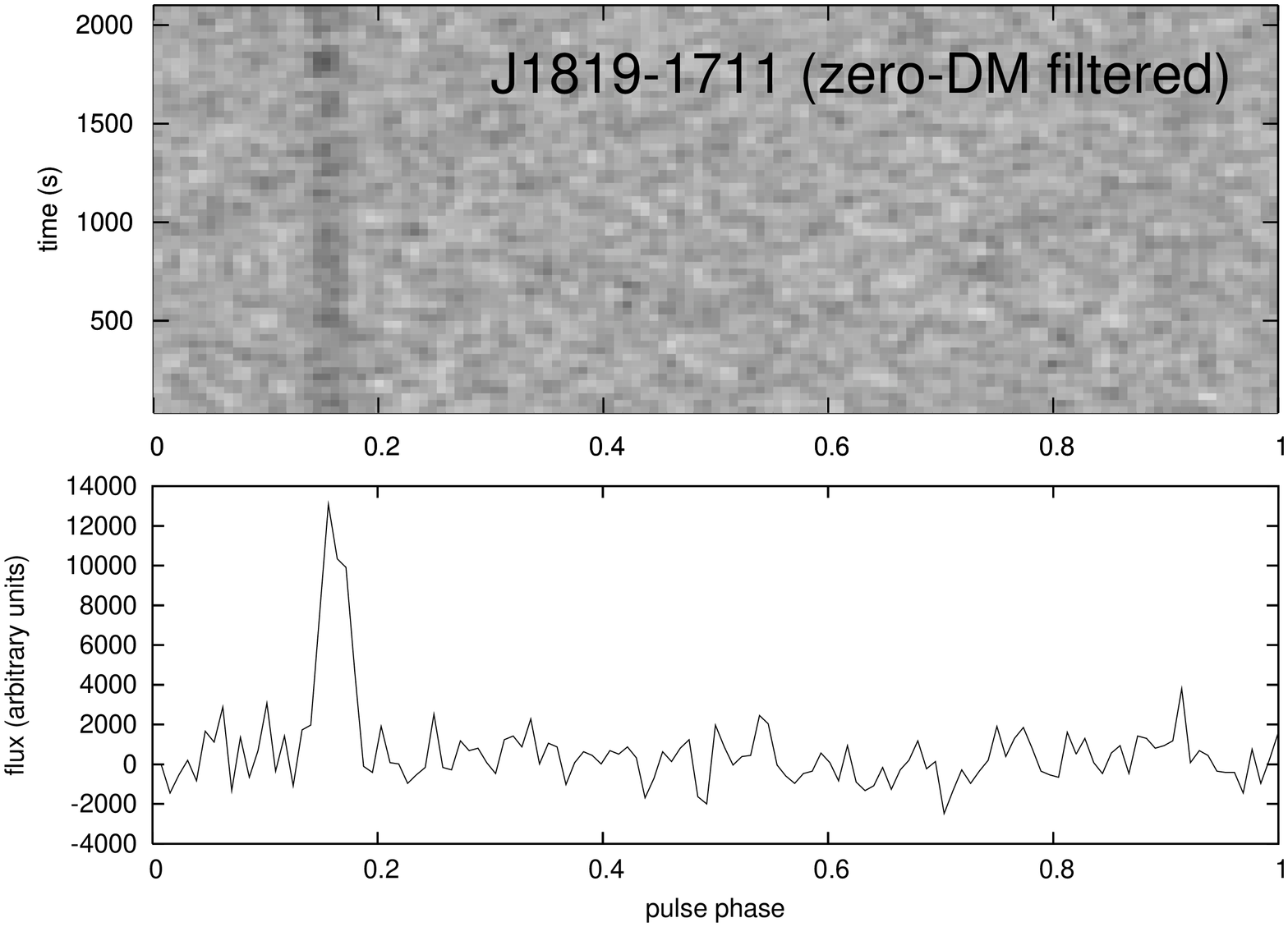}

\includegraphics[angle=-90, scale=0.16]{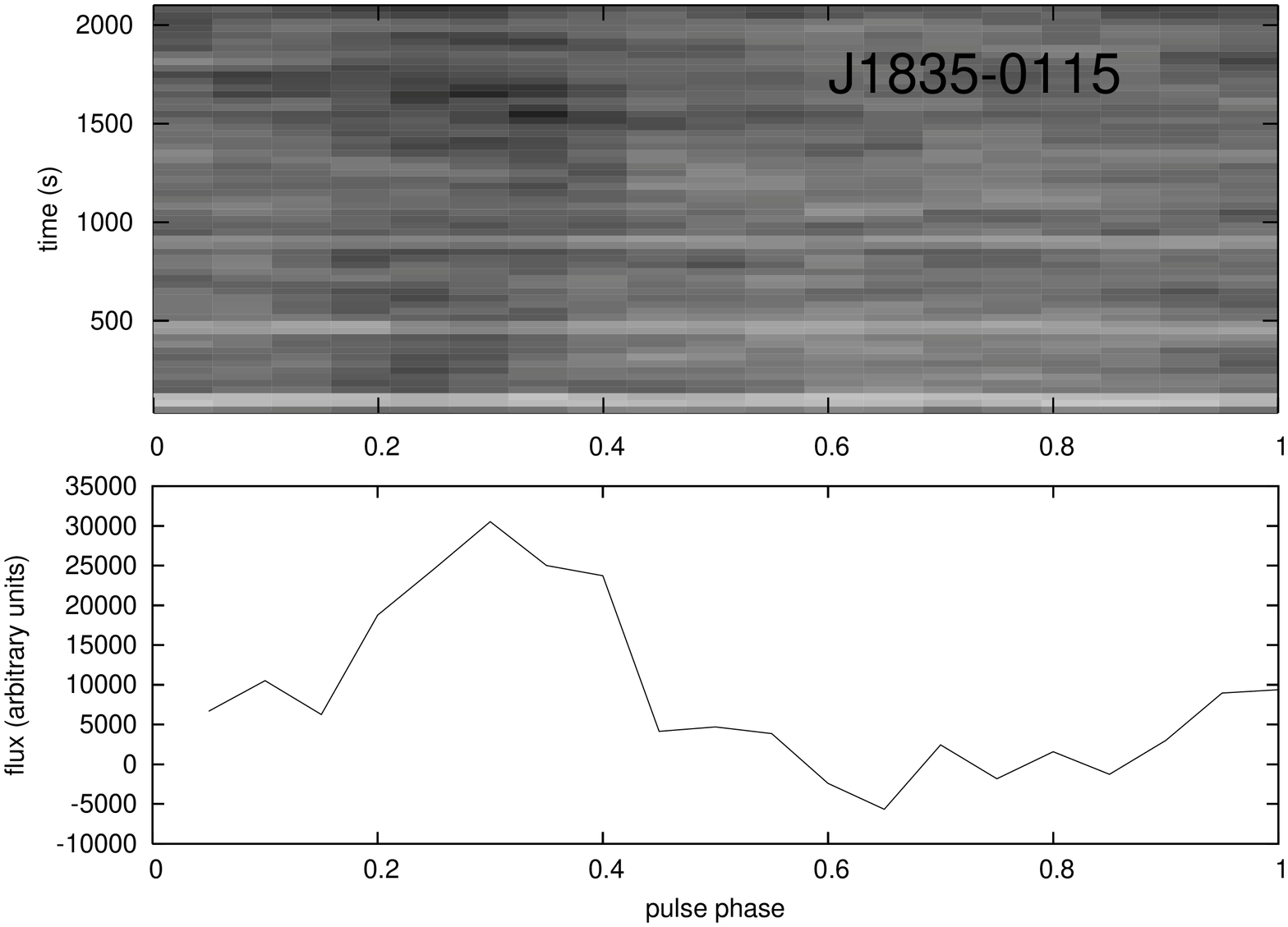}
\includegraphics[angle=-90, scale=0.16]{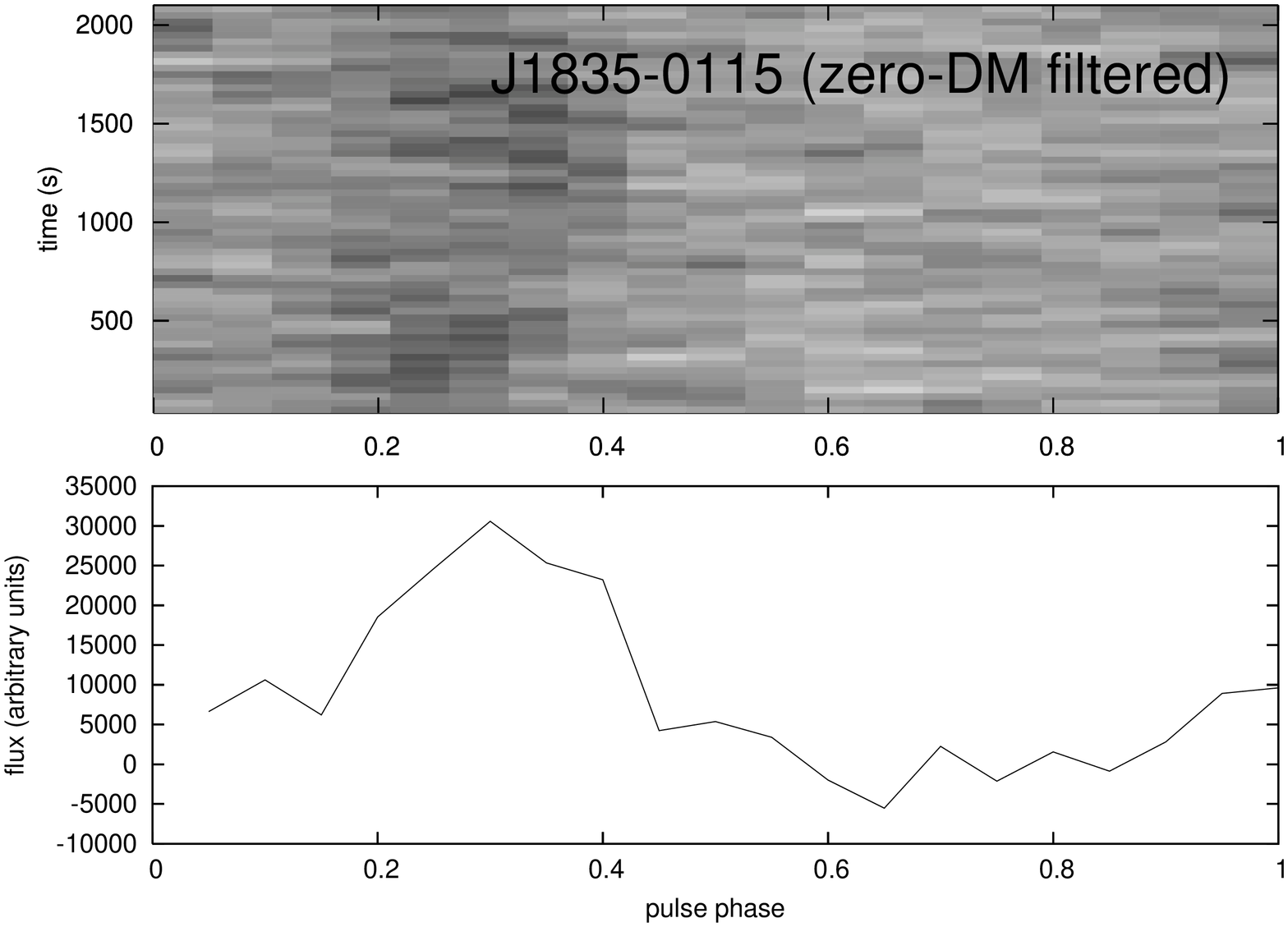}
\caption{\small Folded subintegrations and integrated pulse profiles
for three of the pulsars discovered. For each pulsar we show the
standard analysis on the left and the zero-DM filtered data on the
right.}
\label{fig:newp}
\end{center}
\end{figure}

The zero-DM filter has been incorporated into a new re-analysis of the
PMPS using acceleration searches (Eatough et al. in prep). So far
three new pulsars have been discovered (see Table 1 \&
Figure~\ref{fig:newp}). Figure~\ref{fig:newp} shows folded
subintegrations and integrated pulse profiles for each of the pulsars
in both a standard analysis and after using the zero-DM filter. In all
cases impulsive RFI has been removed.
Although this demonstrates an improvement in the quality of the folded
data, the discovery of these sources was primarily aided by either a
reduction of RFI features in the fluctuation spectra or because wide
spectral filters have not been applied around strong periodic sources
of RFI. Of particular interest is PSR~J1835$-$0115, a 5.1-millisecond
pulsar in a 6-day orbit around a low mass companion. All three pulsars
are now being timed using radio telescopes at either Jodrell Bank or
Parkes.

In addition the zero-DM filter is being applied to our re-analysis of
the PMPS using single-pulse searches (Keane et al. in prep). Here we
present one new source (see Table 1). This source, which has a period
of 1.4s, shows single-pulse diagnostic plots which indicate a strong
resemblance to the newly-discovered class of Rotating RAdio Transients
(RRATs) (McLaughlin et al. 2006)\nocite{mll+06}.  Interestingly, this
object shows pulses from two parts of the pulse rotation period,
separated by about 0.2 sec. We are continuing to monitor this source
using the Parkes radio telescope to further constrain the spin period
and to obtain a period derivative so it can be compared to the values
of period derivative measured for the other RRATs.

\begin{table}
  \begin{center}
    \caption{Preliminary properties of the 4 new sources discovered in
    our re-analyses of the PMPS. The nominal positions quoted are the 
    positions of the centre of the beam in which the source was
    discovered and have errors of less than 7~arcmin, the
    radius of the PMPS beam (Manchester et al. 2001).}
    \begin{scriptsize}
      \begin{tabular}{|l|l|l|r|c}
	\\\hline
	Name & $\alpha$~(J2000) & $\delta$~(J2000) & Period(s) & DM(${\rm cm^{-3}pc})$ \\\hline 
	\hline
	J1724-3549 & 17:24:43.0 & -35:49:18.6 & 1.4085 & 550.0 \\
	J1539-4835 & 15:39:18.7 & -48:35:14.0 & 1.2729 & 136.3 \\
	J1819-1711 & 18:19:50.3 & -17:11:48.0 & 0.3935 & 400.8 \\
	J1835-0115 & 18:35:13.9 & -01:15:17.0 & 0.0051 & 98.1 \\
	\hline
      \end{tabular}
      \end{scriptsize}
  \end{center}
\end{table}

\section{Discussion}
Zero-DM filtering reduces the number of spurious candidates detected
by pulsar search software - both in standard periodicity searches and
in single-pulse searches. A major advantage of zero-DM filtering is
that it eliminates the need to apply arbitrary cleaning techniques to
pulsar search data.  One commonly-used technique is that of clipping
the multi-channel data based upon spikes in the DM=0 time series
(e.g. Hobbs et al 2004)\nocite{hfs+04}. For 1-bit sampled filterbank
data like that in the PMPS, with $n$ channels, the time samples should
have a mean of $n/2$ and a standard deviation of $\sqrt{n}/2$
(e.g. Lorimer \& Kramer 2005)\nocite{lk05}.  Typical clipping
algorithms set each of the $n$ channels to alternate zeros and ones if
the time sample differs from the mean by more than three standard
deviations. Clipping could pose a threat to intense burst-like
dispersed signals that are strong enough to be detected in the low-DM
trials of a pulsar search.  It is precisely these type of transient
events which may reveal exciting new astrophysical phenomena (Lorimer
et al. 2007)\nocite{lbm+07}.  Since zero-DM filtering relies on the
undispersed nature of RFI, rather than its sheer strength, such
signals will not be removed.

The primary RFI excision technique for pulsar periodicity searches up
until now has been frequency domain `zapping'. This method requires
the identification of common signals in the fluctuation spectra from
independent observations at different positions on the sky. Such
signals likely have a local terrestrial origin and enter the receiver
system through the far-out sidelobes of the telescope beam.  Once the
RFI signals have been identified their harmonic sequence is removed
from the fluctuation spectrum.  Typical spectral zapping routines can
remove a few percent of the fluctuation spectrum giving a small chance
that a coincident pulsar frequency or its harmonics could be removed.
The zero-DM filtering process presents a more natural way of removing
such RFI signals without the dangers inherent to zapping.

Although effective at removing RFI in our analyses of the PMPS, we now
note a number of limitations and practicalities of the zero-DM filter.
Firstly the technique will not benefit high-frequency, or small
bandwidth pulsar searches because in this case even signals from
celestial sources will not be highly dispersed.  For example, the
dispersive delay across the entire 576-MHz bandwidth in the Parkes
methanol multi-beam pulsar survey at 6.3 GHz is less than 2 ms for a
source with DM=100 cm$^{-3}$pc (e.g. O'Brien et
al. 2008.\nocite{ojk+08}).  In this case the zero-DM filter would have
a low-frequency cutoff around 230 Hz making detection of all but
millisecond pulsars impossible.

In single-pulse searches, impulsive RFI signals can occasionally be
strong enough to persist even after zero-DM filtering. The remnant RFI
streaks appear as high signal-to-noise features at non-zero
DM. However such remnant RFI streaks will not have the expected drop
off in signal-to-noise ratio as a function of trial DM, nor will they
appear dispersed according to Equation~\ref{eq:disp_t} like a true
celestial source. They are also likely to appear in multiple beams of
a multi-beam receiver. Each of these additional properties can be used
to identify such spurious signals.

An assumption made in the analysis of the implementation of the
zero-DM filter is that of a linear dispersion slope when the true
slope is in fact quadratic. A complete description of the dispersion
slope increases the difficulty of implementing the algorithm, with
little benefit. The effects of this assumption manifest themselves in
phase-folded pulse profiles which have asymmetric dips as shown in
Figure~\ref{fig:J1842} for PSR~J1842+0257, unlike the symmetric dips
we have considered in Figure~\ref{fig:zdm}.

Future pulsar surveys will be performed with higher time and frequency
resolution\footnote{\footnotesize
http://astronomy.swin.edu.au/pulsar/?topic=hlsurvey}.  Fast
dedispersion codes are being developed to cope with the large volume
of data that will be produced (Bailes, private communication). These
dedispersion algorithms operate on 1-bit or 2-bit data before the
samples have been unpacked as floating point numbers. In these
algorithms, calculation of the mean and subsequent subtraction from
each frequency channel sample as in Equation~\ref{eq:zero-DM} may not
be possible at the pre-dedispersion stage.

The factor of two increase in the number of computations imposed by
re-application of the zero-DM filter after every dispersion trial
might not be practical in searches where large numbers of Fast Fourier
Transformations are performed, viz. searches for highly accelerated
binary pulsars (e.g. Eatough et al. in prep). Another practical
consideration is for data sampled with a larger number of bits, i.e. a
large dynamic range. In this scenario any RFI with intensity structure
across the band may `leak' through the zero-DM filter.

We have implemented the zero-DM filter in the {\it Sigproc-4.2} pulsar
analysis software suite and are currently applying the process in a
re-analysis of the PMPS in both acceleration searches (Eatough et
al. in prep) and single-pulse searches (Keane et al. in prep).

\section*{ACKNOWLEDGEMENTS}
This research was partly funded by grants from the Science \&
Technology Facilities Council. Evan Keane acknowledges the support of a
Marie-Curie EST Fellowship with the FP6 Network ``ESTRELA'' under
contract number MEST-CT-2005-19669. The Australia Telescope is funded
by the Commonwealth of Australia for operation as a National Facility
managed by the CSIRO. We would like to thank M. Kramer, B. Stappers,
and C. Jordan for useful discussions and manuscript reading. We also
acknowledge A. Forti and The University of Manchester Particle Physics
group for use of the Tier2 computing facility\footnote {\footnotesize
http://www.hep.manchester.ac.uk/computing/tier2}.

\bibliographystyle{mn2e}

\end{document}